\documentclass[preprint,aps]{revtex4}
\usepackage{graphicx,amsfonts}
\usepackage{dcolumn}
\usepackage{bm}
\begin{document}

\title{Dynamical Breakdown of Symmetry in a (2+1) Dimensional Model Containing 
 the Chern-Simons Field}
\author{Alex G. Dias, M. Gomes and A. J. da Silva}
\affiliation{Instituto de F\'\i sica, Universidade de S\~ao Paulo, \\
Caixa Postal 66318, 05315-970,  S\~ao Paulo, SP Brazil\\
E-mail: alexdias, mgomes, ajsilva @fma.if.usp.br}
\begin{abstract}
We study the vacuum stability of a model of massless scalar and
  fermionic fields minimally coupled to a Chern-Simons field.  The
  classical Lagrangian only involves dimensionless parameters, and the
  model can be thought as a (2+1) dimensional analog of the
  Coleman-Weinberg model.  By calculating the effective potential, we
  show that dynamical symmetry breakdown occurs in the two-loop
  approximation. The vacuum becomes asymmetric and mass generation,
  for the boson and fermion fields takes place. Renormalization group
  arguments are used to clarify some aspects of the solution.
\end{abstract}
\maketitle
\newpage
\section{Introduction}
\label{sec:intro}

Quantum field theories involving the Chern-Simons (CS) field in (2+1) dimensions 
present many peculiar and surprising aspects. Fractional spins, 
exotic statistics and
the existence of massive gauge fields are well known examples
in this direction. Applications of these models are, for example, 
planar Aharonov-Bohm effect, fractional quantum Hall effect and 
surface effects in liquid helium \cite{Hall}.

Physicists working in these theories have
a special concern about their 
phase structure and phase transitions. These properties are investigated 
through the calculation of the renormalization group 
functions and the effective potential. So, renormalization and its 
intermediate step, regularization, 
are central issues to these studies. As the Lagrangian density of the 
CS field involves the Levi-Civita tensor, these models are not easily
extensible outside (2+1) dimensions, making the dimensional 
regularization scheme \cite{tHooft}, a very cumbersome 
procedure. In fact, 
besides the 
extension out of (2+1) dimensions, dimensional regularization for these models 
requires the introduction of an extra regularization term in the Lagrangian.
Of course, a safe escape would be to avoid the
use of regularizations at all, 
like in the BPHZ renormalization program \cite{BPHZ}, but this is also 
a complex procedure due to the zero mass fields involved 
\cite{gomesschroer,albuquerque}.
For these reasons, a simple regularization method called 
dimensional reduction, 
has been largely employed
\cite{Semenoff,pinheiros,advedev} to deal with the CS 
models. It consists in a simplification of dimensional regularization, 
in which all tensor contractions appearing in Feynman graphs, are 
first realized in (2+1) 
dimensions and only the resulting scalar integrals are extended to 
$D=3-\epsilon$ dimensions. But, as stressed by several authors 
\cite{martin}, this method may lead 
to ambiguities, jeopardizing the invariance of the theory under 
gauge transformations.  Nevertheless, given its great simplicity it 
may be worth to use dimensional reduction in practical calculations, 
after validating its 
consistency through the explicit verification of the 
Ward identities related to the gauge symmetry.

In this paper we calculate the effective potential for the theory of a
massless complex scalar field with a sextuple self-interaction and
minimally coupled with a CS field.  Dynamical symmetry breakdown and
mass generation are explicitly verified not in one loop as in the
Coleman-Weinberg (CW) \cite{Coleman73} model, but in the two-loop
approximation.  The Feynman integrals are regularized by dimensional
reduction. The verification that this regularization procedure
preserves the Ward identities, was carried out in previous works
\cite{pinheiros,alex}.

The model is then extended by including a massless fermion minimally
interacting with the CS field, and coupled to the scalar field through
an ``Yukawa'' term \cite{mckeon}.  Up to two loops, the new graphs
that appear in this extended model involve products of two Levi-Civita
tensors at most. As discussed in \cite{martin}, possible differences
between the expressions calculated using dimensional reduction and
dimensional regularization only appear in the product of three and
more Levi-Civita tensors. Therefore, up to the order we are dealing no
breakdown of the Ward identities takes place.  Here again symmetry
breakdown occurs due to two-loop radiative corrections. Curiously, for
a certain relation between the Yukawa and the scalar gauge couplings
the effective potential is insensitive to the fermion gauge
coupling. We also analyze some renormalization group aspects of the
extended model by determining the modifications in the anomalous
dimension of the scalar field and the {\it beta} function associated
with the sextuple self-interaction.

The organization of the paper is as follows: in Sec. \ref{multi} the
effective potential for the scalar/CS model is calculated and the
symmetry breakdown is discussed. In Sec. \ref{fermion} this study
is extended to the scalar/fermion/CS model. In Sec. \ref{group} a
renormalization group analysis of the  model is performed.  In
the Conclusions the comparison with the literature is carried. Three
Appendices present the main integrals necessary in the work and the
calculation of the wave function renormalization for the $\varphi$
field.

\section{The bosonic model}
\label{multi}

In this section we consider the model of a massless complex scalar
field with a sextuple self-interaction and minimally interacting with
a CS field, whose Lagrangian density is

\begin{equation}
{\mathcal{L}}=
\frac{1}{2}\epsilon_{\mu\nu\rho}A^{\mu}\partial^{\nu}
A^{\rho} +(D^{\mu}\varphi)^{\dag}(D_{\mu}\varphi)-
\frac{\nu}{6}(\varphi^{\dag}\varphi)^3, 
\label{1}
\end{equation}

\noindent
where $\nu$ is a positive constant and
$D^{\mu}=\partial^{\mu}-ieA^{\mu}$.  The metric used is
$g^{\mu\nu}=(1,-1,-1)$ and the fully antisymmetric Levi-Civita tensor
$\epsilon_{\mu\nu\rho}$ is normalized as $\epsilon_{012}=1$. It is
worth to stress that all the parameters in this Lagrangian are
dimensionless, what makes it a (2+1) dimensional analog of the well
known CW model in (3+1) dimensions. If instead of the CS, a Maxwell
dynamics for $A_{\nu}$ were used, a dimensional parameter would be
introduced in the Lagrangian and this analogy would be lost.

It is convenient to
decompose the complex scalar field as
$\varphi=(\varphi_1+i\varphi_2)/\sqrt{2}$,
where $\varphi_1$ and $\varphi_2$ are real fields, so that the Lagrangian
(\ref{1}) becomes:

\begin{eqnarray}
{\mathcal{L}}&=&\frac{1}{2}(\partial^{\mu}\varphi_i)^2+
e\epsilon^{ij}A_{\mu}\varphi_j\partial^{\mu}\varphi_i+
\frac{e^2}{2}A^{\mu}A_{\mu}\varphi_i\varphi_i-\frac{\nu}{48}(\varphi_i
\varphi_i)^3 
+\frac{1}{2}\epsilon_{\mu\nu\rho}A^{\mu}\partial^{\nu}A^{\rho},
\label{2}
\end{eqnarray}

\noindent
where $\epsilon^{12}=- \epsilon^{21}=1$ and the summation convention is
implied, with $i$ and $j$ running over 1 and 2.
We shall work in a 't Hooft
gauge, which has the advantage of eliminating non
diagonal terms in
the free propagators. The gauge fixing term is taken as

\begin{eqnarray}
{\mathcal{L}}_{GF}=-\frac{1}{2\xi}(\partial^{\mu}A_{\mu}-\xi
e\epsilon^{ij}\varphi_iu_j)^2,
\label{3}
\end{eqnarray}

\noindent
and the corresponding Faddeev-Popov Lagrangian is given by

\begin{eqnarray}
{\mathcal{L}}_{FP}=-c^{\dagger}(\partial^2 +\xi
e^2\epsilon^{ij}\varphi_iu_j)c,
\label{4}
\end{eqnarray}

\noindent
where $u_i$ is a constant to be chosen later.  To calculate the
effective potential we follow the functional method \cite{jackiw74}.
The first step is to consider the action,
$\hat{I}(\varphi_i,\chi;\phi_i)=\int
d^3x{\hat{\mathcal{L}}}(\varphi_i,\chi;\phi_i)$, obtained by shifting
the scalar fields by a constant: $\varphi_i\rightarrow\varphi_i+\phi_i
$ in the original Lagrangian (\ref{2}) and subtracting the terms which
are either independent or linear in the fields, i. e.,

\begin{eqnarray}
\hat{I}(\varphi_i,\chi;\phi_i)=I(\varphi_i+\phi_i,\chi)-I(\phi_i,0)-\int d^3x
\frac{\delta I}{\delta\varphi_i}(\phi_i,0)\,\varphi_i- 
\int d^3x\frac{\delta I}{\delta \chi}(\phi_i,0)\,\chi, 
\label{5}
\end{eqnarray}
\noindent
where $\phi$ is the constant expectation value of $\varphi$ and $\chi$ 
represents the fields $A_{\mu}$, $c$ and $c^{\dagger}$.

By choosing $u=\phi$, bilinear terms in $A_{\mu}$ and $\varphi_i$ are
eliminated from the resulting ${\hat{\mathcal{L}}}$, which becomes

\begin{eqnarray}
\hat{\mathcal{L}}&=&\frac{1}{2}A_{\mu} [  \epsilon_{\mu \nu \rho}
\partial_{\rho}+ m_3 g^{\mu \nu}+\frac{1}{\xi}\partial^{\mu}\partial^{\nu}]
A_{\nu} +c^{\dagger} ( -\partial^2-m_3\xi )c\nonumber\\
&+&\frac{1}{2}\varphi_i [\delta_{ij} (-\partial^2-m_1^2-m_3 \xi )+ (m_3\xi
-4m_1^2 )\hat{\phi}_i\hat{\phi}_j ]\varphi_j+\nonumber\\
&-&e A^{\mu}\epsilon_{ij}\varphi_i\partial_{\mu}\varphi_j
+e^2 \phi\cdot \varphi A^2-e^2\xi \phi \cdot\varphi c^{\dag}c\nonumber\\
&-&\frac{\nu}{6}(\phi \cdot \varphi)^3
-\frac{\nu}{4}\phi^2 \phi \cdot \varphi\varphi^2+\frac{e^2}{2}A^2\varphi^2
-\frac{\nu}{16}\phi^2\varphi^4\nonumber\\
&-&\frac{\nu}{4}(\phi \cdot \varphi)^2 \varphi^2
-\frac{\nu}{8}\phi \cdot \varphi \varphi^4-\frac{\nu}{48} \varphi^6.
\label{6}
\end{eqnarray}

\noindent
where $m_1^2=\nu \phi^4/8$, $m_3=e^2\phi^2$,
$\hat{\phi}_i=\phi_i/\sqrt{\phi^2}$ and $\phi \cdot \varphi$, $\phi^2$
and $\varphi^2$ stand respectively for $\phi_i \varphi_i$,
$\phi_i\phi_i$ and $\varphi_i \varphi_i$. For later use we will also
define $m_2^2=5 m_1^2$.

The effective potential is given by 

\begin{eqnarray}
V=\frac{\nu}{48}\phi^6-\frac{i}2\int \frac{d^3\hspace{0.1 cm}k}{(2\pi)^3}\ln
[\det\hspace{0.1 cm}(i\Delta_{\alpha\beta}^{-1}(k,\phi_i))]+ i <0|\,
T\exp i\!\! \int d^3 x\hat {\cal  L}_{int}|0>.
\label{7}
\end{eqnarray}

\noindent
The first term in (\ref{7}) is the tree approximation as can be read
from (\ref{2}).  The second term is the one-loop correction and the
$i\Delta_{\alpha\beta}^{-1}(k,\phi_i)$ is the coefficient matrix of
the quadratic terms of $\hat{\mathcal{L}}$. The third term is the sum
of the two and more loops vacuum diagrams, calculated from
$\hat{\mathcal{L}}$; it summarizes the infinity sum of all higher
order loop diagrams with any number of external scalar lines
$\varphi=\phi$, gotten from the original Lagrangian (\ref{2})
\cite{Coleman73}.
  
The one-loop effective potential is obtained using the inverse propagators 
from (\ref{6}), that are

\begin{eqnarray}
&&i \Delta^{-1}_{ij}(k)=\delta_{ij}(k^2-m_{_1}^2-\xi 
m_{_3})+\hat{\phi}_i\hat{\phi}_j(-4m_{_1}^2+\xi m_{_3}),
\label{8a}\\\nonumber \\
&&i \Delta^{-1}_{\mu\nu}(k)=-i\epsilon_{\mu\rho\nu}k^\rho+
m_{_3}g_{\mu\nu}-\frac{1}{\xi}k_\mu k_\nu,
\label{8b}\\\nonumber\\
&&i \Delta^{-1}_{ghost}(k)=k^2-\xi m_{_3},
\label{8c}
\end{eqnarray}

\noindent
for the scalar, gauge and ghost fields respectively. As said in the
Introduction,  we  employ the dimensional reduction
method in which the integrals are promoted to $D=3-\epsilon$ and a mass
scale $\mu $ is introduced to keep the dimensions of the relevant
quantities unchanged ($\int d^3k \rightarrow \mu^{\epsilon} \int d^{3-\epsilon}k$). Thus, using (\ref{8a})-(\ref{8c}), we obtain that the one-loop contribution
is:

\begin{eqnarray}
V^{B(1) } =-\frac{1}{12\pi}[m_{_1}^3(1+5^{\frac{3}{2}})+m_{_3}^3]
=-\frac{1}{12\pi}[\nu^{3/2}\frac{1+5^{3/2}}{16\sqrt{2}}+e^6]
\phi^6. \label{8d}
\end{eqnarray}

\noindent
Here and in what follows we are retaining only the contributions that
do not vanish as $\epsilon$ tends to zero. We have also chosen
to work in the Landau  limit ($\xi \rightarrow0$) where the ghosts decouple  and do not contribute to the
potential. 

As a consequence of the dimensional reduction regularization
in (2+1) dimensions, in one loop, no infinity appears.  Up to one
loop, the effective potential is given by: $V(\phi)=
\frac{\nu}{48}\phi^6 +V^{B(1)}(\phi)+\frac{C}{48} \phi^6$, where $C$ is
a convenient finite counterterm. It does not have any non trivial
$\phi\neq 0$ minimum, and dynamical breakdown of symmetry does not
occur.  The symmetry breakdown in the CW model in (3+1) dimensions is
made possible by the induction of a term of the form $\phi^4
\ln\phi$.  In the analogous calculation in (2+1) dimensions,
the dimension of the phase space precludes the induction of a similar
term (actually $\phi^6\ln\phi$). The only effect, of the one-loop calculation, is the change in
the coefficient of the $\phi^6$ term of the classical potential, and
so symmetry breakdown does not happen.  Thus, to pursue a possible symmetry
breakdown we will study the two-loop approximation.

The propagators (in the Landau limit)
for the $A^\mu$   and $\varphi$  fields obtained from the shifted
Lagrangian are:

\begin{eqnarray}
\Delta_{\mu\nu}(k)=-\frac{\epsilon_{\mu\nu\rho}k^\rho}{k^2-m_{_3}^2}-i\frac{m_{_3}}{k^2-m_{_3}^2}( g_{\mu\,\nu}-\frac{k_{\mu}\,k_{\nu}}{k^2}),
\label{9b}
\end{eqnarray}

\begin{eqnarray}
\Delta_{ij}(k)=i\,[\,\frac{1}{k^2-m_{_1}^2}(\delta_{ij}-\hat{\phi}_i\hat{\phi}
_j)+\frac{1}{k^2-m_{_2}^2}\hat{\phi}_i\hat{\phi}_j\,].
\label{9a}
\end{eqnarray}

The interaction vertices are given by
(we consider only those vertices which contribute to the two-loop calculation):

\begin{eqnarray}
\mbox{Quadrilinear $A_\mu A_\nu \varphi_i \varphi_j$ vertex} \qquad&\leftrightarrow&\qquad \frac{i}2 e^2 \delta_{ij}g_{\mu\nu} \\  
\mbox{Quadrilinear $\varphi_i\varphi_j\varphi_k\varphi_l$ vertex}
\qquad&\leftrightarrow&\qquad -i \frac{\nu\phi^2}{8}(\frac12\delta_{ij} + 
2 \hat \phi_i\hat \phi_j)\delta_{kl}\\
\mbox{Trilinear $\varphi_i \varphi_k \varphi_j$ vertex} \qquad&\leftrightarrow&\qquad -i\frac{\nu}4\phi^3 \hat \phi_i(\delta_{jk} + 
\frac23\hat\phi_j\hat\phi_k)\\
\mbox{Trilinear $\varphi_i A_\mu A_\nu$ vertex} \qquad&\leftrightarrow&\qquad ie^2\phi \hat\phi_i g_{\mu\nu}\\
\mbox{Trilinear $\varphi_i \varphi_j A_\mu$ vertex} \qquad&\leftrightarrow&\qquad -\frac{e}{2} \epsilon_{ij} (p+q)_\mu
\end{eqnarray}

The two-loop contributions to the effective potential are drawn in
Fig.~\ref{Fig2}.  The corresponding analytic expressions are listed in
the Appendix \ref{a2}. In the dimensional reduction approach the Lorentz
indices are contracted in three dimensions, and only after all the
tensor simplifications are done, the resulting scalar integrals are
promoted to $D=3-\epsilon$ dimensions.  The calculations, tedious but
straightforward, are carried by using the formulas presented in the
Appendices.  $V^{B(2)}_{\ref{Fig2}a}$ and $V^{B(2)}_{\ref{Fig2}b}$, which correspond to the
diagrams in Figs. \ref{Fig2}$a$ and \ref{Fig2}$b$ respectively, are
convergent, since they are given by products of non overlapping
one-loop integrals. The other three diagrams have divergences
proportional to $\phi^6$, in accordance with the renormalizability of the
model.  Summarizing, the results are

\begin{eqnarray}
V^{B(2)}_{\ref{Fig2}a}=\frac{\nu }{2^7\pi^2}\,\phi^2\,(\frac{3}{2}{m_{1}}^2+3{m_{1}}{m_{2}}
+\frac{11}{2}{m_{2}}^2),
\end{eqnarray}

\begin{eqnarray}
V^{B(2)}_{\ref{Fig2}b}=\frac{e^2}{2^4\pi^2}({m_{1}}{m_{3}}^2+{m_{2}}{m_{3}}^2),
\end{eqnarray}

\begin{eqnarray}
V^{B(2)}_{\ref{Fig2}c} =-\frac{1}{2^5}\,\nu^{2}\,[2\,I(m_{1}, \,m_{2}, \,m_{1}) + {\displaystyle \frac {50}{3}} \,I(m_{2}, \,m_{2}, \,m_{2})]\,\phi^6,
\end{eqnarray}

\begin{eqnarray}
V^{B(2)}_{\ref{Fig2}d}&=& \frac {1}{2} e^{2}\,m_{3}\,
\Big{[}-\frac {1}{2^3\pi^2} \, (\,
m_{1}\,m_{3} + \,m_{2}\,m_{3})  +2(\,m_{1}^{2}+ \,m_{2}^{2} )
\,I(m_{2}, \,m_{1}, \,m_{3})\nonumber \\ 
&-&  \frac{2}{m_3^2}\, [K(m_{2}, \,m_{1}, \,m_{3})
+ K(m_{1}, \,m_{2}, \,
m_{3}) - K(m_{2}, \,m_{1}, \,0) - K(m_{1}, \,m_{2}, \,0)\,]\,\Big{]} 
\end{eqnarray}

\noindent
and

\begin{eqnarray}
V^{B(2)}_{\ref{Fig2}e}&=&e^{4}\,\phi^{2} \Big{[}\frac {3}{2^4\pi^2} \,m_{3}\,m_{2}
  - 2\,J(m_{3}, \,m_{3}, \,m_{2}) + J(0, \,m_{3}, \,m_{2})\nonumber \\ 
&-& 6\,m_{3}^{2}\,I(m_{3}, \,m_{3}, \,m_{2})
+ 3\,m_{3}^{2}\,I(0, \,m_{3}, \,m_{2})   
-\frac{1}{m_3^2}\,[\,K(m_{3}, \,m_{3}, \,m_{2})\nonumber \\ 
&-& K(m_{3}, \,0, \,m_{2}) - K(0, \,m_{3}, \,m_{2}) +  K(0, \,0, \,m_{2})\,]\,\Big{]}.
\end{eqnarray}

\noindent
After using the results of the Appendix \ref{a1}, where the functions
$I$, $J$ and $K$ are defined,  and for convenience introducing 
 $I_{div}=\frac{1}{\epsilon}-\gamma+\ln\hspace{0.1 cm}4\pi+1$, we have

\begin{eqnarray}
V^{B(2)}_{\ref{Fig2}a}={\displaystyle \frac {1}{2^{10}\pi ^{2}}}
\,\nu^{2}\,(29 + 3\,\sqrt{5})\,\phi^6,\label{10a}
\end{eqnarray}

\begin{eqnarray}
V^{B(2)}_{\ref{Fig2}b}={\displaystyle \frac {1}{2^5\sqrt{2} \pi ^{2}}} \,e^{6}\,\sqrt{\nu }\,(1 + \sqrt{5})\,\phi^6,\label{10b}
\end{eqnarray}

\begin{eqnarray}
V^{B(2)}_{\ref{Fig2}c}&=&{\displaystyle \frac {1}{3\,2^9 \pi ^{2}}} \,{\displaystyle 
{\nu ^{2}\,\phi^6\,\Big{[}6\,\ln(2 + \sqrt{5}) + 56\,
\ln(\sqrt{\frac{\nu}8 }) + 50\ln3 + 25\,\ln5\Big{]}}}\,\nonumber\\&+& 
{\displaystyle \frac {7}{3\,2^6\pi ^{2}}}\,\nu^2\, \phi^{6}\,
\Big{[}\,{\displaystyle 
 {\,\ln({\displaystyle \frac {\phi^{2}}{\mu }} )-\frac{1}{2}I_{div}}}\,\Big{]},\label{10c}
\end{eqnarray}

\begin{eqnarray}
V^{B(2)}_{\ref{Fig2}d}&=&{\displaystyle \frac {1}{2^5\pi ^{2}}}\,\phi ^{6}\,
\Big{[} - \sqrt{\frac{\nu}8 }\,e^{6} - \,\sqrt{\frac{5\nu}{8} }\,e^{6} +( 3\,e^{8}
-\frac{3}{2}\nu \,e^{4}  + \frac{1}{4}\,\nu ^{2})\ln(\sqrt{\frac{\nu}8 } + 
\,\sqrt{\frac{5\nu}8 } + e^{2}) \nonumber \\ 
&+& \frac{\sqrt{5}}{8}\,\nu \,e^{4}-\frac{1}{4}\sqrt{\frac52}\,\nu ^{3/2}\,e^{2} + 
\frac{1}{4 \sqrt{2}}\,\nu ^{3/2}\,e^{2} -\frac{1}{4}\,\nu ^{2}\,\ln(1 + 
\sqrt{5}) - \frac{1}{4}\,
\nu ^{2}\,\ln(\sqrt{\frac{\nu}8 }) \nonumber\\ 
 &+&  2\,\pi ^{2}\,\nu 
^{2}) \Big{]}+{\displaystyle \frac {1}{2^5\pi ^{2}}} \phi^{6}\,\Big{[}\,
{\displaystyle 
 {\,\ln({\displaystyle \frac {\phi^{2}}{\mu }
} )-\frac{1}{2}I_{div}\,\Big{]}\,( -\frac{3}{2}\,\nu \,e^{4} + \,e^{8})}} 
\label{10d}
\end{eqnarray}

\noindent
and

\begin{eqnarray}
V^{B(2)}_{\ref{Fig2}e}&=& - {\displaystyle \frac {1}{2^7 \pi ^{2}}}\,\phi^{
6}\,\Big{[} - 20\,e^{6}\,\,\sqrt{\frac{5\nu}8 } + 10\,\nu \,{\rm 
ln}(2\,e^{2} + \,\sqrt{\frac{5\nu}8 })\,e^{4} + 12\,e^{8} \nonumber \\
&+& 20\,\ln(e^{2} + \,\sqrt{\frac{5\nu}8 })\,e^{8} \mbox{} - 5\,
\ln(e^{2} + \,\sqrt{\frac{5\nu}8 })\,e^{4}\,\nu  
- 48\,e^{8}\,\ln(2\,e^{2} + \,\sqrt{\frac{5\nu}8 })\nonumber \\ 
&-& \frac{5}{4}\,e^{4}\,\nu  -\frac{5^2}{2^5}\,\ln(2\,e^{2} + \,
\sqrt{\frac{5\nu}8 })\,\nu^{2}+\frac{5^2}{2^4}\,\ln(e^{2} + \,\sqrt{
\frac{5\nu}8 })\,
\nu^{2} \nonumber\\
&-&  \frac{5^2}{2^6}\,\nu^{2}\,\ln(5) - \frac{5^2}{2^5}\,\nu^{2}\,
\ln(\sqrt{\frac{\nu}8 })\Big{]} 
+ {\displaystyle \frac {1}{2^5\pi ^{2}}} \phi^6\,\Big{[}\,{\displaystyle  {\ln({\displaystyle \frac {\phi^{2}}{\mu }} )-\frac{1}{2}I_{div}\,\Big{]}\,( - \frac{5}{4}\,e^{4}\,
\nu + 7\,e^{8})}}. \label{10e}
\end{eqnarray}

\noindent
Collecting all these two-loop contributions we obtain:
\begin{eqnarray}
V^{B(2)}=X_b(e,\nu,\epsilon)\,\phi^6 
+Z_b(e,\nu)\,\phi^6\ln(\frac{\phi}{\sqrt{\mu}}),
\label{11}
\end{eqnarray}

\noindent
with $X_b(e,\nu, \epsilon)$ standing for the sum of all coefficients of $\phi^6$ in Eqs. 
(\ref{10a}-\ref{10e}) and

\begin{eqnarray}
Z_b(e,\nu)=\frac{1}{8\pi^2}(4e^8 -\frac{11}{8}e^4\nu+\frac{7}{12}\nu^2).
\end{eqnarray}
\noindent

From (\ref{7}), (\ref{8d}) and (\ref{11}) we can write the regularized
effective potential up to two loops (with a counterterm C included)
as:

\begin{eqnarray}
V^{B}_{reg}=Z_b \,\phi^6 \, \ln(\,\phi\,e^{Y_b/Z_b}/\sqrt{\mu}\,).
\label{12}
\end{eqnarray}

\noindent
where $Y_b(e,\nu)$ is the constant

\begin{equation}
Y_b(e,\nu, \epsilon)=\frac{\nu}{48} -\frac{1+5\sqrt{5}}{3\,2^{13/2}\pi} 
\nu^{3/2}
-\frac{e^6}{12\pi}+X_b(e,\nu,\epsilon)+\frac{C}{48}.
\end{equation}

\noindent
The parameters $\mu$ and $C$ can be eliminated by imposing the condition

\begin{eqnarray}
\frac{dV^{B}_{ren}}{d\phi}\Big{|}_{\phi=v}&=&0,
\end{eqnarray}

\noindent
where $v$  is an arbitrary non null parameter. The resulting potential is

\begin{eqnarray}
V^{B}_{ren}=Z_b \,\phi^6 \, (\ln \frac{\phi}{v}\,-\frac{1}{6})
\label{13}
\end{eqnarray}

\noindent
and $\phi=v$ is a local minimum (vacuum) of this effective potential, if the 
generated squared mass of the scalar field

\begin{equation}
m_{\phi}^2\equiv \frac{d^2V^{B}_{ren}}{d\phi^2}\Big{|}_{\phi=v}=6\,Z_b\,v^4,
\end{equation}

\noindent
is positive, what means $Z_b>0$. We will choose $Z_b$ through the condition

\begin{eqnarray}
\frac{d^6V^{B}_{ren}}{d\phi^6}\Big{|}_{\phi=v}=15\nu\equiv 
\frac{d^6V^{B}_{tree}}{d\phi^6},\label{pec1}
\end{eqnarray}

\noindent
what implies in

\begin{equation}
\nu=\frac{548}{5}Z_b=\frac{137}{10{\pi}^2}(4 e^8-\frac{11}{8}e^4\nu+
\frac{7}{12}{\nu}^2)\approx 1.39\,(4 e^8-\frac{11}{8}e^4\nu+
\frac{7}{12}{\nu}^2).
\label{13a}
\end{equation}

\noindent
The solution of this equation for both $\nu$ and $e^2$ in the perturbative 
regime ($\nu$ and $e^2$ $<<1$) is given by

\begin{eqnarray}
\nu \approx \frac{274}{5\pi^2}e^8+{\cal{O}}(e^{12}),\label{pec32}
\end{eqnarray}

\noindent
and in the leading approximation the effective potential and the generated
squared mass result to be

\begin{eqnarray}
V^{B}_{ren}(\phi)=\frac{e^8}{2\pi^2}\,\phi^6\,(\,\ln\frac{\phi}{v}
-\frac{1}{6}\,),
\end{eqnarray}
\noindent

\begin{equation}
m_{\phi}^2=\frac{d^2V^{B}_{ren}}{d\phi^2}\Big{|}_{\phi=v}=\frac{3}{\,
\pi^2}\,e^8 v^4>0.
\end{equation}

\noindent
The ratio of this mass to the induced squared  mass  of $A_{\mu}$, that is
$m_{A}^2\approx m_3^2(\phi=v) =e^4 v^4$, gives: $m_{\phi}^2/m_{A}^2\approx
3 e^4/\pi^2$.

Summarizing, the CW mechanism is operative due to the two-loop
radiative corrections, that is, in this approximation a non trivial
vacuum is induced and masses for the boson and CS fields are
generated.  Similarly to what happens in four dimensions, no symmetry
breakdown occurs in the model of a single scalar field in 
self-interaction.  Indeed, by making $e=0$ in (\ref{13a}) we get $\nu=120
\pi^2/959\approx1.23$ and so, out of the perturbative regime of
validity of the calculations.

The two-loop effective potential for the scalar/CS model was
previously calculated in \cite{hosotani}, using dimensional
regularization. As shown by the authors, in that regularization
scheme, the extension out of (2+1) dimensions is not enough to
regularize the model; it becomes also necessary, to introduce in the
Lagrangian, an extra regularization term, a Maxwell like term:
$-\frac{a}{4}F^{\mu \nu}F_{\mu \nu}$, depending on a parameter $a$.
No trouble is found when making the regulators $\epsilon$ and $a$ to
go to zero after the renormalization, in the calculation of the
effective potential. This is not  true in the calculation of the
renormalization group parameters (see discussion in Sec. \ref{group}).

\section{Adding fermions to the model}
\label{fermion}

Many interesting phenomena in planar physics, 
which have the CS model as an effective
theory, also involve fermion particles. 
Since the advent of the CS 
field theory \cite{des87} a vast literature (see \cite{dun99} and the
references therein) 
on the subject has appeared, but the effect of fermions on 
the effective potential for a model like (\ref{2}), does not seem
to have been studied. In this Section we will 
extend the model  (\ref{2}), by including a Dirac fermionic field 
interacting with all the other fields. This is done by adding to (\ref{2}) 
the following Lagrangian density:

\begin{eqnarray}
{\mathcal{L}}_{Dirac}=i\bar{\psi}\gamma^\mu(\partial_\mu-iqA_\mu)\psi-\frac{\alpha}{2}\varphi_i\varphi_i\bar{\psi}\psi,
\label{14}
\end{eqnarray}

\noindent
where $\psi$ is a two-component massless Dirac field that represents a
particle and its antiparticle with the same spin projection. The
$\gamma^\mu$ matrices were chosen as
$(\gamma^0,\gamma^1,\gamma^2)=(\sigma^3,i\sigma^1,i\sigma^2)$, where
$\sigma^{a}$ are the Pauli matrices. Besides the minimal interaction
with the CS field, the fermion (charge $q$) also couples to the
scalar field through an ``Yukawa'' term with the coupling constant
$\alpha$. As the parameters of the purely bosonic model, these new
coupling constants, $\alpha$ and $q$, are also dimensionless.

After shifting the scalar field: $\varphi \rightarrow \varphi+\phi$, as in 
Section \ref{multi} we get

\begin{eqnarray} 
\hat{{\mathcal{L}}}_{Dirac}=i\bar{\psi}\gamma^\mu \partial_\mu \psi 
+q\bar{\psi}\gamma^{\mu} A_\mu\psi-\frac{\alpha \phi^2}{2}\bar{\psi}\psi-\frac{\alpha}{2}\varphi_i\varphi_i
\bar{\psi}\psi -\alpha \phi\cdot \varphi\, \bar{\psi}\psi.
\label{15}
\end{eqnarray}

\noindent
The fermion propagator  is $S(p)=i/(\gamma^\mu p_\mu-m_{_4})$ with
$m_{4}=\alpha \phi^2/2$, and the interaction vertices  are given by  
\begin{eqnarray}
\mbox{Trilinear $A_\mu  \bar \psi \psi$ vertex} 
\qquad&\leftrightarrow&\qquad iq\gamma_\mu \\  
\mbox{Trilinear $\varphi_i\bar \psi \psi$ vertex}
\qquad&\leftrightarrow&\qquad -i \alpha\phi \hat\phi_i\\
\mbox{Quadrilinear $\varphi_i \varphi_j \bar \psi\psi$ vertex} \qquad&\leftrightarrow&\qquad -i\frac{\alpha}2\delta_{ij}
\end{eqnarray}

The one-loop contribution to the effective potential
is 

\begin{eqnarray}
V^{F(1)}=i\int \frac{d^3k}{(2\,\pi)^3}\,\ln\,\det(\gamma^{\mu}p_{\mu}-m_4)=
\frac{1}{6\pi}m_{_4}^3={\displaystyle \frac {\alpha^3}{48\pi}} \,
{\displaystyle {\phi^{6}}},
\end{eqnarray}

\noindent
and the new two-loop contributions are represented by the 
diagrams in Fig.~\ref{Fig4}. The results are (see Appendix \ref{a2})

\begin{eqnarray}
V^{F(2)}_{2a}=-\frac{\alpha}{4\pi^2}\Big{(}m_{_4}^2m_{_1}+m_{_4}^2m_{_2}\Big{)},
\end{eqnarray}

\begin{eqnarray}
V^{F(2)}_{2b}=\phi^2\frac{\alpha^2}{2}\Big{(} -\frac{1}{8\pi^2}\,m_4\,m_2 +J(m_4, 
\,m_2, \,m_4)+4m_4^2\,I(m_4, \,m_2, \,m_4)\Big{)},
\end{eqnarray}

\noindent
and

\begin{eqnarray}
V^{F(2)}_{2c}=&&2 q^2\Big{[} 
-\frac{1}{16\pi^2}\,m_4^3+(m_3^2\,m_4-m_4^2\,m_3)I(m_4, 
\,m_3, \,m_4) +\frac{m_3}{2}J(m_4, \,m_3, \,m_4)\nonumber\\
&&-\frac{1}{m_3}(K(m_4, \,0, \,m_4)-K(m_4, \,m_3, 
\,,m_4))\Big{]}.
\end{eqnarray}

\noindent
With the aid of the formulas in the Appendices we have

\begin{eqnarray}
V^{F(2)}_{2a}= - \frac {\alpha^3}{16\pi ^2} \,\phi^6 \,
\Big(\sqrt{\frac\nu{8}} +\,\sqrt{5\frac\nu{8} }  \Big),
\end{eqnarray}

\begin{eqnarray}
V^{F(2)}_{2b}=&-&\frac {\alpha^2}{128\pi ^2} \,
\phi^6\,\Big[4\,\alpha \,
\,\sqrt{\frac{5\nu}{8} } - \frac {20\,\nu}{8} \,\ln(\alpha  + 
\,\sqrt{\frac{5\nu}{8} }) - \alpha ^2 + 4\,\alpha ^2\,{\rm 
ln}(\alpha  + \,\sqrt{\frac{5\nu}{8}})\Big ]\nonumber \\
&-& \frac {1}{32\pi^2} \, 
 \phi^6\,\Big[\,{\displaystyle 
 {\,\ln({\displaystyle \frac {\phi^{2}}{\mu }} \,)-\frac{1}{2}I_{div}}}\,\Big{]}\,( - \frac {5\,\nu}{8}\,\alpha ^{2}  + \alpha ^{4}) ,
\end{eqnarray}

\noindent
and

\begin{eqnarray}
V^{F(2)}_{2c}=&& \!\frac {1}{128\pi ^{2}} q^{2}\,\phi^{6} 
\Big[ - 2\,\alpha ^{3} + 4\,e^{6}\,\ln(\alpha  + e^{2}) + e
^{2}\,\alpha ^{2} - 8\,e^{4}\,\alpha \,\ln(\alpha  + e^{2})
  + 4\,e^{2}\,\alpha ^{2}\,\ln(\alpha  + e^{2}) \nonumber\\
&&\!- 4\,\alpha 
\,e^{4}\Big]+ \frac {q^2}{32\pi ^{2}} \, \phi^{6}\,\Big{[}\,{\displaystyle 
 {\,\ln({\displaystyle \frac {\phi^{2}}{\mu }} )-\frac{1}{2}I_{div}}}\,\Big{]}
\,(e^{6} - 2\,e^{4}\,\alpha  
+ e^{2}\,\alpha ^{2}). 
\end{eqnarray}

Thus the fermion sector contributes to the 
effective potential with 
\begin{eqnarray}
V^{F}=X_f \phi^6+Z_f \phi^6 \ln(\frac{\phi}{\sqrt{\mu}}),
\label{15b}
\end{eqnarray}

\noindent
where $X_f$ is the sum of the coefficients of the terms $\phi^6$ in 
the previous expressions and 
$Z_f$ is the sum of the coefficients of the terms 
$\phi^6 \ln(\phi/\sqrt{\mu})$ which results to be
\begin{eqnarray}
Z_f(q,\alpha,\,\nu)=\frac {1}{16\pi ^{2}}( \frac{5\,\nu}{8}\,
\alpha ^{2}  - \alpha 
^{4}+q^{2}e^6 - 2\,q^{2}e^4\,\alpha  + q^{2}e^2\,\alpha ^{2}).
\label{zf}
\end{eqnarray}

The complete effective potential is the sum of (\ref{12}) 
and (\ref{15b}) and is given by
\begin{eqnarray}
V_{reg}=Z\,\phi^6\,\ln(\frac{\phi\,e^{Y/Z}}{\sqrt{\mu}}),
\label{Vtot}
\end{eqnarray}
\noindent
where $Y=Y_b+X_f$ and 
\begin{eqnarray}
Z=Z_b+Z_f=\frac{1}{8\pi^2}[4e^8 -\frac{11}{8}e^4\nu+\frac{7}{12}\nu^2 
+\frac{5}{16}\,\nu\,\alpha ^2-\frac12\alpha ^4+\frac12 q^2 e^2\,
(e^2-\alpha) ^2].
\label{15a}
\end{eqnarray}

\noindent
After using the vanishing of the first derivative of $V(\phi)$ in $\phi=v$, 
it can be written as
\begin{eqnarray}
V_{ren}(\phi)=Z\,{\phi}^6[\ln(\frac{\phi}{v})-\frac{1}{6}].
\label{16}
\end{eqnarray}

\noindent
As in the pure bosonic model, the positivity of the induced squared
mass requires that $Z>0$, and as there, we choose to fix it through
the condition (\ref{pec1}), what leads  to an equation similar to (\ref{13a})
for $Z$.  In the perturbative regime (that is, $\nu,\, e,\, q$ and $\alpha$
$<<1$), this equation implies that the second, the third and the
fourth terms on the right hand side of (\ref{15a}) are infinitesimals of higher
order and can be dropped, leaving the equation
\begin{eqnarray}
\nu = \frac{548}{5}Z \approx \frac{137}{10\pi^2}[4e^8 +\frac12 q^{2}e^2(e^2
-\alpha)^2 -\frac12\alpha^4].
\label{17}
\end{eqnarray}

\noindent
Dynamical symmetry breakdown and mass generation occur if the
condition $4e^8+\frac12 q^{2}e^2(e^2-\alpha)^2>\frac12\alpha^4$ is
satisfied, what is true for a continuum of values of the coupling
constants, their magnitudes chosen as $\nu \sim e^8,\,q^8,\,\alpha^4<<1$.
Some particular cases are worth being mentioned: 1. choosing
$\alpha=e^2$, we still have the solution $\nu\approx
\frac{137}{10\pi^2}(\frac72 e^8)$, but the result becomes independent of the
charge of the fermion field. 2. dropping the Yukawa interaction (by making
$\alpha=0$), we get $\nu\approx \frac{137}{10\pi^2}(4e^8 +\frac{q^{2}e^6}2)$,
showing that the fermion indirect interaction with the scalar field
(mediated by its interaction with the CS field) reinforces the
symmetry breakdown.  3. for $e=q=0$, that is, for the model of a boson
in self-interaction, and interacting with a fermion field through an
Yukawa term, we see from (\ref{17}) that no symmetry breakdown is
possible; this result is in agreement with the conclusions of
\cite{mckeon} in which (besides other possibilities) a similar model
with a Dirac two-component fermionic field and a real scalar field was
considered.
4. symmetry breakdown does not occur if we only take $e=0$ leaving the
scalar field indirectly interact with the CS through its coupling 
with the fermion field.

\section{Renormalization group analysis}
\label{group}

In this section some renormalization group aspects of the previous solution 
for the effective potential will be discussed. 
The regularized effective potential satisfies the renormalization group
equation 
\begin{eqnarray}
\Big{[}\mu\frac{\partial}{\partial\mu}+\beta_\nu\frac{\partial}{\partial\nu} +\beta_e\frac{\partial}{\partial e} +\beta_q\frac{\partial}{\partial q} +\beta_\alpha\frac{\partial}{\partial\alpha} -\gamma_\varphi\phi\frac{\partial}{\partial\phi}\Big {]}V_{reg}(\mu,\epsilon,\nu,e,q,\alpha,\phi)=0,   
\label{rg}
\end{eqnarray}

\noindent
where
\begin{eqnarray}
\beta_\nu=\mu\frac{d\nu}{d\mu}, \hspace{0.2 cm}\beta_e=\mu\frac{de}{d\mu}, \hspace{0.2 cm}\beta_q=\mu\frac{dq}{d\mu}, \hspace{0.2 cm}\beta_\alpha=\mu
\frac{d\alpha}{d\mu}, \hspace{0.2 cm}
\end{eqnarray}
\noindent
and
\begin{eqnarray}
\gamma_\varphi=\frac{\mu}{2Z_\varphi}\frac{dZ_\varphi}{d\mu},
\end{eqnarray}

\noindent
are respectively the coupling constants beta functions and the scalar field
anomalous dimension, and  $Z_\varphi$ is the wave function
renormalization constant of the scalar field. 

As mentioned before dimensional reduction in 2+1 dimensions automatically
removes the divergences of the one-loop graphs.  Thus
 non-trivial  renormalization group parameters are obtained only when the two-loop approximation is
considered. By applying the operator inside the bracket of
(\ref{rg}) to the regularized expression (\ref{Vtot}),
we get the following result 
\begin{eqnarray}
\Big{(}-\frac{Z}{2}+\frac{\beta_\nu^{(2)}}{48}-\frac{6\nu}{48}\gamma_\varphi^{(2)}\Big{)}\phi^6=0,
\end{eqnarray}

\noindent
where the superscript $(2)$ indicates the loop order of the corresponding
function. Thus
\begin{eqnarray}
\beta_\nu^{(2)}=24Z+6\nu\gamma_\varphi^{(2)},\label{pec2}
\end{eqnarray}
\noindent
with $Z$ given by Eq. (\ref{15a}).

To obtain $\gamma_\varphi^{(2)}$ we have to calculate $Z_\varphi$ up
to two loops. This can be done by considering the model in its
symmetric phase where all fields are massless. The Feynman diagrams
which give non-trivial contributions to $Z_\varphi$ are given in
Fig.~\ref{Fig5}. Three of them were computed in ref. \cite{pinheiros}
and the results are quoted in Appendix \ref{a3} where the calculation
of the remaining ones is also summarized. From (\ref{last}) we get
\begin{eqnarray}
\gamma_\varphi^{(2)}=-\frac{1}{48\pi^2}[7e^4+2(q^2e^2-\alpha^2)].\label{pec3}
\end{eqnarray}

By replacing this expression in Eq. (\ref{pec2})  we finally have
\begin{eqnarray}
\beta_\nu^{(2)}=\frac{1}{\pi^2}\Big{[}\frac{7}{4}\nu^2+e^4(-5\nu+12e^4)+
\frac{\alpha^2}{2}(\frac{17}{8}\nu-3\alpha^2)+\frac{q^2e^2}{2}((\alpha-e^2)^2
-\nu)\Big{]}.
\end{eqnarray}

For the bosonic model ($q=0=\alpha$) this yields: 
\begin{eqnarray}
\gamma_\varphi^{(2)}&=&-\frac{7}{48\pi^2}e^4,\\
\beta_\nu^{(2)}&=&\frac{7}{4\pi^2}(\nu^2-\frac{20}{7}\nu e^4+
\frac {48}{7} e^8),\label{rg2}
\end{eqnarray}
\noindent
confirming the results obtained in \cite{albuquerque} by using soft
BPHZ \cite{gomesschroer} and also in \cite{pinheiros} by using
dimensional reduction. These results are also in qualitative agreement
with that of \cite{advedev} but, differently of what happens with the
effective potential, disagree with those of \cite{hosotani} (which in
our notation are $\gamma_\varphi=0$ and $\beta_\nu=\frac{7}{4\pi^2}
\nu^2$) where dimensional regularization with minimal subtraction was
used. However, as discussed in \cite{hosotani}, dimensional
regularization with minimal subtraction is not perturbatively
consistent for the pure CS model (some of the $\beta$'s functions
diverge when the regulating Maxwell term is removed).  On general
grounds, the authors of \cite{hosotani} argue that $\gamma_\varphi$
and $\beta_\nu$ should explicitly depend on $e^2$, a characteristic
present in our results but not in the ones obtained by the use of the
dimensional regularization with minimal subtraction scheme.

An interesting property of the $\beta_{\nu}$ function given in
(\ref{rg2}) is the fact that it does not vanish for $\nu=0$.  From the
knowledge that $e$ does not change with the renormalization scale (in
[7] we showed that $\beta^{(2)}_{e}=0$) we conclude that
$\gamma_\varphi$ is a constant and so
$\phi(\mu)=(\mu_0/\mu)^{\gamma_{\varphi}} \phi(\mu_0)$ and the
equation $\mu d\nu/d\mu=\beta_{\nu}$ can immediately be integrated
resulting in:
\begin{equation}
\frac{2\nu-ce^4}{e^4b}=\Big[\frac{
\frac{2\nu_0-ce^4}{e^4b}+\tan(ae^4\ln(\mu/\mu_0)
)}{1-\frac{2\nu_0-ce^4}{e^4b}\tan(ae^4\ln(\mu/\mu_0))}\Big],\label{pec30}\\
\end{equation}
where $a=\sqrt{59}/ (2\pi^2)\approx 0.39$, $b=4\sqrt{59}/7\approx
4.39$ and $c=20/7\approx 2.86$.

The effective potential is invariant under renormalization group
transformations, i.e. $V(\phi,\nu,e,\mu)=V(\phi_0,\nu_0,e,\mu_0)$.
Therefore, as the above solution for $\nu$ is regular at $\nu_0=0$
($\beta_\nu\not =0$ for $\nu=0$), then by conveniently choosing
$\mu_0$ we get $V(\phi,\nu,e,\mu)=V(\phi_0,0,e,\mu_0)$. This means
that the effective potential in presence of the sextuple
self-interaction can be obtained from the simpler model in which the
boson only interacts with the CS field.

From the above expressions for $V$ and $\phi$ we also have 
\begin{equation}
\frac{dV}{d\phi}(\phi,\nu,e,\mu) =\Big(\frac{\mu}{\mu_0}\Big)^{\gamma_\varphi}
 \frac{dV}{d\phi_0}(\phi_0,\nu_0,e,\mu_0) ,\label{pec31}
\end{equation}
so that starting at $\phi_0=v$ and $\nu_0\approx \frac{274}{5 \pi^2}
e^4$, which implies that $dV/d\phi=0$ for a certain value of $\mu_0$,
one can go to the values of $\nu$ specified by Eq. (\ref{pec30}) and
$\phi(\mu)=(\mu_0/\mu)^{\gamma_{\varphi}} v$ which due to
Eq. (\ref{pec31}) also corresponds to $dV/d\phi=0$. This 
shows that the condition (\ref{pec32}) for dynamical symmetry
breakdown can be relaxed; the only restriction to get a symmetry
breakdown is, in fact, that all the coupling constants be small.

A similar analysis for the complete model, i.e. with the inclusion of
the fermion fields, would require a lot more calculations ($\beta_q,
\beta_{\alpha}$, etc) and we do not pursue it here. However, some
observations are in order: 1. differently from the other couplings,
the Yukawa coupling increases the anomalous dimension of $\phi$ (see
(\ref{pec3})). 2. as for the bosonic model, $\beta_{\nu}$ does not
vanish for $\nu=0$, and as above the effective potential can be get
from the simpler model with $\nu=0$.  3. as in the purely bosonic
model, it is also expected that the constraint (\ref{17}) on $\nu$ can
be relaxed, the only restriction being that $\nu<<1$ (as it must be
for the others coupling constants).

\section{Conclusions}
\label{sec:con}

In this paper we calculated the effective potential, up to two
loops, for a (2+1) dimensional model composed of an interacting
massless scalar field, a massless fermion field interacting with the
scalar field through an Yukawa term, and a CS gauge field, minimally
coupled to the scalar and the fermionic fields.  As the CW model in
(3+1) dimensions, it only involves massless parameters, and is
classically invariant under scale transformations, what makes it a
possible candidate for dynamical symmetry breakdown. As we verified,
dynamical symmetry actually takes place, but differently from the
original model of CW, in which this effect already manifests in
one-loop corrections, here it only shows up starting in two loops. 
For particular values of the couplings our
effective potential coincides with those found 
in the literature: by discarding the fermion contribution, it agrees
with the previous calculations of \cite{hosotani} for the same model
without fermions; if instead we drop the contributions involving the
CS field, it agrees with the results of \cite{mckeon} for a model
without the gauge field. 

We also calculated the renormalization group functions $\beta_\nu$ and $\gamma_\varphi$ for the extended model. For the pure bosonic (sub) model they agree
with our previous calculation using other regularization/renormalization
techniques but disagree with the results of \cite{hosotani}. It would be
interesting to compute the other renormalization group functions of the
extended model.

\section{Acknowledgments}

This work was partially supported by Funda\c c\~ao de Amparo a
Pesquisa do Estado de S\~ao Paulo (FAPESP) and Conselho Nacional de 
Desenvolvimento Cient\'{\i}fico e Tecnol\'ogico (CNPq).

\newpage
\begin{appendix}

\section{Useful Integrals.}
\label{a1}

\vspace{0.1 cm}

At two loops the following integrals \cite{hosotani} appear ($d^D p \equiv \mu^\epsilon d^{3-\epsilon} p$):

\begin{eqnarray}
I(m_{_1},m_{_2},m_{_3})=\int\frac{{d^Dp}\hspace{0.1 
cm}{d^Dq}}{(2\pi)^{2D}}
\frac{1}{((p+q)^2-m^{2}_{_1})(q^2-m^2_{_2})(p^2-m^2_{_3})}\nonumber\\[10pt]
=\frac{1}{32\pi^2}(\frac{1}{\epsilon}-\gamma+\ln\hspace{0.1 
cm}4\pi+1)-\frac{1}{16\pi^2}\ln[\frac{m_{_1}+m_{_2}+m_{_3}}{\mu}],
\end{eqnarray}

\begin{eqnarray}
K(m_{_1},m_{_2},m_{_3})&=&\int\frac{{d^Dp}\hspace{0.1 
cm}{d^Dq}}{(2\pi)^{2D}}\hspace{0.1 cm} 
\frac{(p \cdot q)^2}{((p+q)^2-m^2_{_1})(q^2-m^2_{_2})(p^2-m^2_{_3})}\\
&=&\frac{1}{64\pi^2}[m^2_{_1}(m_{_1}m_{_2}+m_{_1}m_{_3}-m_{_2}m_{_3})-m_{_1}(3m^3_{_2}+3m^3_{_3}+m^2_{_2}m_{_3}\nonumber\\
&+&m_{_2}m^2_{_3})
+(m^2_{_2}+m^2_{_3})m_{_2}m_{_3}]+\frac{1}{4}(m^2_{_1}-m^2_{_2}-m^2_{_3})^2
\,\,I(m_{_1},m_{_2},m_{_3})\nonumber
\end{eqnarray}

\noindent
and

\begin{eqnarray}
J(m_{_1},m_{_2},m_{_3})&=&\int\frac{{d^Dp}\hspace{0.1 
cm}{d^Dq}}{(2\pi)^{2D}}\hspace{0.1 cm} 
\frac{2(p\cdot q)}{((p+q)^2-m^2_{_1})(q^2-m^2_{_2})(p^2-m^2_{_3})}\\
& &=(m^2_{_1}-m^2_{_2}-m^2_{_3})I(m_{_1},m_{_2},m_{_3})-\frac{1}{16\pi^2}
[m_{_2}m_{_3}-m_{_1}m_{_2}-m_{_1}m_{_3}]. \nonumber
\end{eqnarray}

\section{Two-loop diagrams}
\label{a2}

The analytic expressions for the two-loop vacuum diagrams contributing
to the effective potential shown in  Figs. \ref{Fig2} and \ref{Fig4} are: 

\begin{eqnarray}
V^{B(2)}_{\ref{Fig2}a}&=&\frac{\nu}{8}\phi^2(\frac{\delta_{ij}}{2}+2\hat\phi_i\hat\phi_j)
\delta_{kl}
\int\frac{d^{^D}p\hspace{0.2 cm}d^{^D}q}{(2\pi)^{2D}}\nonumber\\
&&\{ \Delta_{ij}(p)\Delta_{kl}(q)+\Delta_{jl}(p)\Delta_{ki}(q)+
\Delta_{jk}(p)\Delta_{li}(q)\} 
\label{Vq1}
\end{eqnarray}

\begin{eqnarray}
V^{B(2)}_{\ref{Fig2}b}=-\frac{e^2}{2}\delta_{ij}g_{\mu\nu}\int
\frac{d^{^D}p\hspace{0.2 cm}d^{^D}q}{(2\pi)^{2D}}\Delta_{ij}(p)
\Delta^{\mu\nu}(q)
\label{Vq2}
\end{eqnarray}

\begin{eqnarray}
V^{B(2)}_{\ref{Fig2}c}&=&-i\frac{\nu^2\phi^6}{32}\hat\phi_i
(\delta_{jk}+\frac{2}{3}\hat\phi_j\hat\phi_k)\hat\phi_l(\delta_{mn}+\frac{2}{3}
\hat\phi_m\hat\phi_n)\nonumber\\ 
&\times&\int\frac{d^{^D}p\hspace{0.2 
cm}d^{^D}q}{(2\pi)^{2D}}\Big{\{}\Delta_{il}(q)[\Delta_{jm}(p)
\Delta_{kn}(p+q)+\Delta_{jn}(p)\Delta_{km}(p+q)]\nonumber\\
&+&\Delta_{im}(q)[\Delta_{jl}(p)\Delta_{kn}(p+q)+\Delta_{jn}(p)
\Delta_{kl}(p+q)]\nonumber\\ 
&+&\Delta_{in}(q)[\Delta_{jm}(p)\Delta_{kl}(p+q)+\Delta_{jl}(p)
\Delta_{km}(p+q)]\Big{\}}
\label{Vt1}
\end{eqnarray}

\begin{eqnarray}
V^{B(2)}_{\ref{Fig2}d}&=&i\frac{e^2}{4} 
\int\frac{d^{^D}p\hspace{0.2 
cm}d^{^D}q}{(2\pi)^{2D}}(2q+p)^\mu(2q+p)^\nu\epsilon^{ij}\epsilon^{kl}
\Delta_{\nu\mu}(p)\nonumber\\
&&\Big [ \Delta_{li}(q)\Delta_{jk}(p+q)
-\Delta_{ki}(q)\Delta_{jl}(p+q)\Big ]
\label{Vt2}
\end{eqnarray}

\begin{eqnarray}
V^{B(2)}_{\ref{Fig2}e}=-ie^4\phi^2 
\int\frac{d^{^D}p\hspace{0.2 
cm}d^{^D}q}{(2\pi)^{2D}}\hat\phi_i\hat\phi_j g_{\mu\nu}g_{\sigma\rho}
\Delta^{\sigma\mu}(p+q)\Delta_{ji}(p)\Delta^{\nu\rho}(q)
\label{Vt3}
\end{eqnarray}
\noindent

\begin{eqnarray}
V^{F(2)}_{\ref{Fig4}a}=-\frac{\alpha}{2}\int\frac{d^{^D}p
\hspace{0.2 cm}d^{^D}q}{(2\pi)^{2D}}tr[S(p)\Delta_{ii}(q)]
\label{Vq3}
\end{eqnarray}
\noindent

\begin{eqnarray}
V^{F(2)}_{\ref{Fig4}b}=i\frac{({\alpha}\phi)^2}{2}\hat\phi_i\hat\phi_j
\int\frac{d^{^D}p\hspace{0.2 cm}d^{^D}q}{(2\pi)^{2D}}tr[S(p)S(p+q)
\Delta_{ij}(q)]\label{Vt4}
\end{eqnarray}
\noindent

\begin{eqnarray}
V^{F(2)}_{\ref{Fig4}c}=i\frac{{q}^2}{2}\int
\frac{d^{^D}p\hspace{0.2 
cm}d^{^D}q}{(2\pi)^{2D}}tr[S(p)\gamma^\mu S(p+q)\gamma^\nu\Delta_{\nu\mu}(q)]
\label{Vt5}
\end{eqnarray}
\noindent

\section{Wave function renormalization of  the $\varphi$ field up to  two loops}
\label{a3}

As known in three dimensions the use of dimensional reduction removes
the divergences of one-loop graphs. Therefore the $\varphi$ field
renormalization constant only receives nontrivial contributions
starting in two loops.

In two loops the nonvanishing contributions to $Z_\varphi$ come from
the graphs depicted in Fig.~\ref{Fig5}. The first three graphs, Figs. \ref{Fig5}$(a-c)$
were calculated in \cite{pinheiros} and here we just quote the result

\begin{equation}
Z^{(2)}_{\varphi(a-c)}=\frac{7 e^4}{48\pi^2} \frac{1}{\epsilon}.
\end{equation}
 
\noindent
For the remaining graphs, a direct calculation furnishes:

\begin{equation}
\mbox{Graph in Fig. {\ref{Fig5}$d$}} = -i 4e^2\epsilon^{\mu\nu\rho}\epsilon^{\alpha\beta\sigma}\int\frac{d^D k }{(2\pi)^{D}}
\frac{ k_\rho  k_\sigma 
p_\nu p_\alpha \pi_{\mu\beta}(k)}{(k^2)^2(p+k)^2}
=-i \frac{e^2q^2 p^2}{24\pi^2} \frac1\epsilon+ 
\mbox{finite terms},
\end{equation}

\noindent
where $\pi_{\mu\beta}$, given by the upper loop in Fig.~\ref{Fig5}$d$,  is the
fermion contribution to the polarization tensor

\begin{equation}
\pi_{\mu\beta}(k) = -\frac{q^2}{16}  (g_{\mu\beta} k^2 - k_\mu k_\beta)
\frac{1}{(k^2)^{2-D/2}}
\end{equation}

\noindent
and 

\begin{eqnarray}
\mbox{Graph in Fig. {\ref{Fig5}$e$}} &=& \alpha^2\int\frac{d^D k d^D q}{(2\pi)^{2D}}
tr[S( q) S(p+q+k)] \Delta(k)\nonumber\\&=&- 2i\alpha^2 \int\frac{d^D k d^D q}{(2\pi)^{2D}}\,
\frac{q^2+ q\cdot (k+p)}{q^2k^2(p+k+q)^2} \nonumber\\
&=& i \frac{\alpha^2 p^2}{96\pi^2} \frac1\epsilon+ 
\mbox{finite terms}.
\end{eqnarray}

\noindent
Thus up to two loops  the total wave function renormalization of
the $\varphi$ field is

\begin{equation}\label{last}
 Z^{}_{\varphi}= 1+ \frac{1}{24\pi^2}(\frac{7e^4}2+e^2 q^2-\frac{\alpha^2}{4})
\frac1\epsilon.
\end{equation}
\end{appendix}

\newpage
\begin{figure}[ht]
\includegraphics{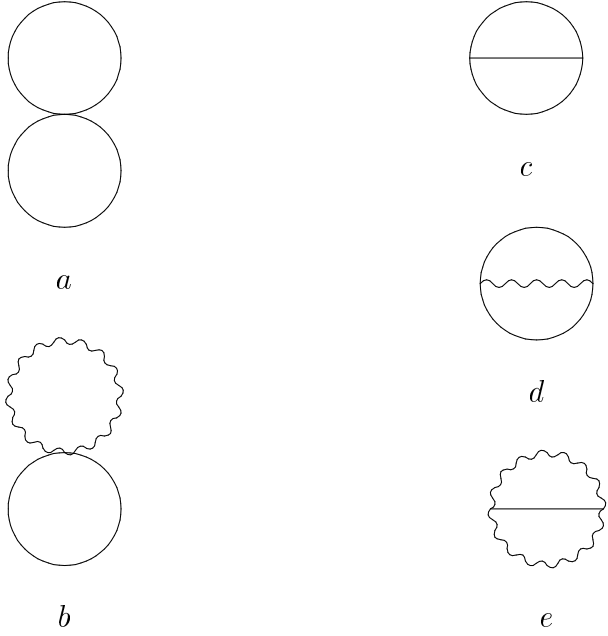}
\caption{Diagrams contributing to the two-loop approximation for the effective 
potential. Continuous and wavy lines represent respectively the $\varphi$ and $A_\mu$ field
propagators.}
\label{Fig2}
\end{figure}

\begin{figure}[ht]
\includegraphics{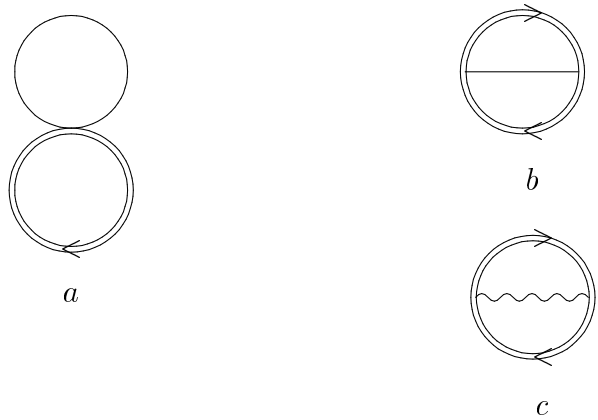}
\caption{ Additional diagrams contributing to the two-loop approximation to the effective 
potential when fermions are present. Double lines represents the fermion field
propagator}
\label{Fig4}
\end{figure}

\begin{figure}[ht]
\includegraphics{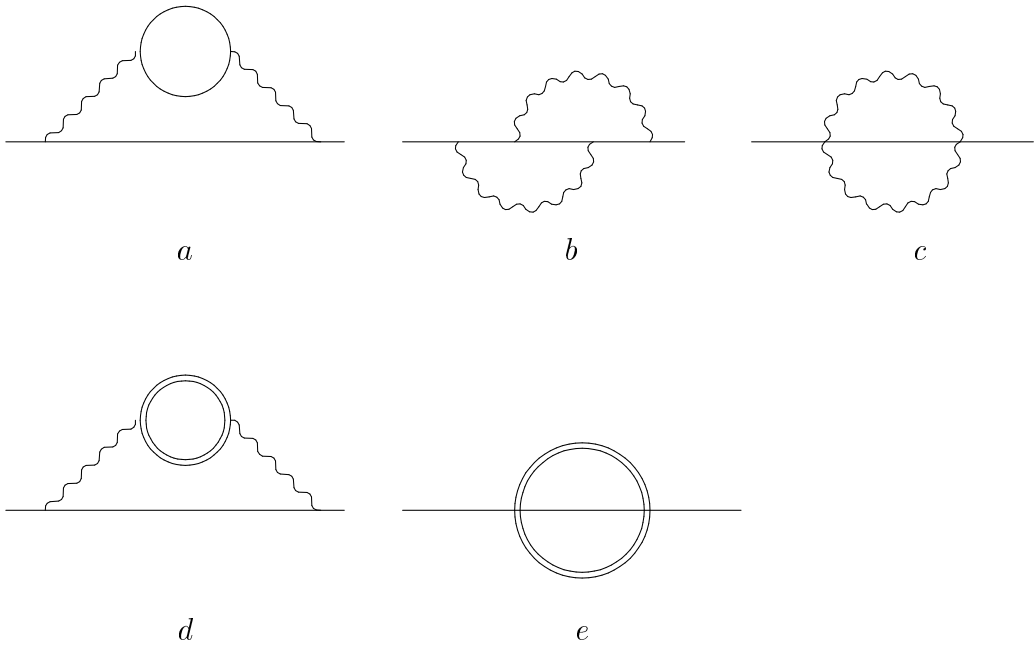}
\caption{Divergent diagrams contributing to the two-loop approximation to 
the two point function of the scalar field.}
\label{Fig5}
\end{figure}
\end{document}